\documentclass[preprint,showpacs,preprintnumbers,superscriptaddress,
amsmath,showkeys,amssymb,aps,floatfix]{revtex4}

\usepackage{graphicx}
\usepackage{dcolumn}
\usepackage{bm}
\usepackage{longtable}
\usepackage{epsfig}
\usepackage{times}
\begin{document}

%\begin{frontmatter}

\title{Nuclear excitation by electron capture followed by fast x-ray emission}

%\author[MPI]{Adriana~P\'alffy \corauthref{cor}},
%\corauth[cor]{Corresponding author}
%\ead{Palffy@mpi-hd.mpg.de}
%\author[MPI]{Zolt\'an~Harman},
%\author[GSI]{Christophor~Kozhuharov},
%\author[GSI]{Carsten~Brandau},
%\author[MPI]{Christoph~H.~Keitel},
%\author[Gi]{Werner~Scheid},
%\author[GSI]{Thomas St\"ohlker}

%\address[MPI]{Max-Planck-Institut f\"ur Kernphysik, Saupfercheckweg~1, 
%69117 Heidelberg, Germany}
%\address[GSI]{Gesellschaft f\"ur Schwerionenforschung (GSI), Planckstr.~1, 64291 Darmstadt, Germany}
%\address[Gi]{Institut f\"ur Theoretische Physik, Justus-Liebig-Universit\"at Giessen, Heinrich-Buff-Ring~16, 35392 
%Giessen, Germany}

\author{Adriana~P\'alffy}\email[Corresponding author, postal address: Max-Planck-Institut f\"ur Kernphysik, Saupfercheckweg~1, 69117 Heidelberg, Germany, telephone +49 (0)6221 516153, fax +49 (0)6221 516152, electronic address: ]{Palffy@mpi-hd.mpg.de}

\author{Zolt\'an~Harman}
\affiliation{Max-Planck-Institut f\"ur Kernphysik, Saupfercheckweg~1,
69117 Heidelberg, Germany}
\author{Christophor~Kozhuharov}
\author{Carsten~Brandau}
\affiliation{Gesellschaft f\"ur Schwerionenforschung (GSI), Planckstr.~1, 64291
Darmstadt, Germany}
\author{Christoph~H.~Keitel}
\affiliation{Max-Planck-Institut f\"ur Kernphysik, Saupfercheckweg~1,
69117 Heidelberg, Germany}
\author{Werner~Scheid}
\affiliation{Institut f\"ur Theoretische Physik, Justus-Liebig-Universit\"at Giessen, Heinrich-Buff-Ring~16, 35392
Giessen, Germany}
\author{Thomas St\"ohlker}
\affiliation{Gesellschaft f\"ur Schwerionenforschung (GSI), Planckstr.~1, 64291
Darmstadt, Germany}
\affiliation{Physikalisches Institut, Ruprecht-Karls-Universit\"at  Heidelberg, Philosophenweg 12,
69120 Heidelberg, Germany}

\date{\today}

\begin{abstract}
The resonance strength of the two-step process of nuclear excitation by electron capture followed by $\gamma$~decay of the nucleus can be significantly increased in highly charged ions if the resonant capture proceeds via an excited electronic state with subsequent fast x-ray emission.
For fully ionized $^{238}_{92}\mathrm{U}$ and $^{232}_{90}\mathrm{Th}$, the \mbox{x-ray} decay  stabilizes the system against internal conversion of the captured electron, with an increase of both  nuclear lifetimes and  resonance strengths of up to two orders of magnitude compared with the case when occupied atomic orbitals prevent the x-ray de-excitation.  Applications of this effect to the measurement of the not yet experimentally observed nuclear excitation by electron capture and to dense astrophysical plasmas are discussed.
\end{abstract}



\maketitle

The population and lifetime of nuclear excited states can be affected by the electronic
shells, particularly in the case of nuclear processes which directly involve electrons such as 
nuclear electron capture (EC) and internal conversion (IC). In few-electron highly charged ions, the strongly-bound inner-shell electrons can significantly influence the nuclear decay, leading to effects on IC and EC that are contrary to intuition. 
For instance,  decay measurements  of the 14.4~keV M\"ossbauer level in $^{57}_{26}\mathrm{Fe}$ in one- and two-electron ions have shown that the nuclear lifetime is about $20\%$ shorter in H-like $\mathrm{Fe}^{25+}$ ions   than  in He-like  $\mathrm{Fe}^{24+}$ ions or in the neutral atom \cite{Philips}.
 Similar experimental results have been obtained for EC rates in H-, He-like and neutral $^{140}_{59}\mathrm{Pr}$, where the nuclear lifetime of the one-electron $\mathrm{Pr}^{58+}$ ion is shorter than the ones of the corresponding two- or many-electron cases \cite{Yuri}.
 The time-reversed processes of IC and EC, namely, nuclear excitation by electron capture (NEEC) and bound $\beta$ decay, respectively, require the presence of vacancies in the atomic shell. The opening of the new bound $\beta$ decay channel in highly charged ions considerably  influences the lifetime of unstable levels in nuclei \cite{Takahashi1,Takahashi2}.  As a spectacular example, the lifetime of the   ground state $^{187}_{75}\mathrm{Re}$ decreases by more than nine orders of magnitude  from 42~Gyr for the neutral atom to 32.9 yr for bare ions as a consequence of new bound $\beta$ decay branches to the ground and excited states of the $^{187}_{76}\mathrm{Os}$ daughter \cite{Bosch2,Bosch1}. The case  of $^{187}_{75}\mathrm{Re}$ is particularly interesting in astrophysical context, since  it affects the accuracy of the $^{187}\mathrm{Re}$-$^{187}\mathrm{Os}$ cosmochronometer \cite{Bosch1}. The motivation in investigating the behavior of nuclei in highly charged ions is thus related to potential interests in nuclear astrophysics, studies of nuclear decay properties and tests of the  relativistic description of atomic inner-shell processes. 

\begin{figure*}[h]
\begin{center}
\includegraphics[width=0.9 \textwidth]{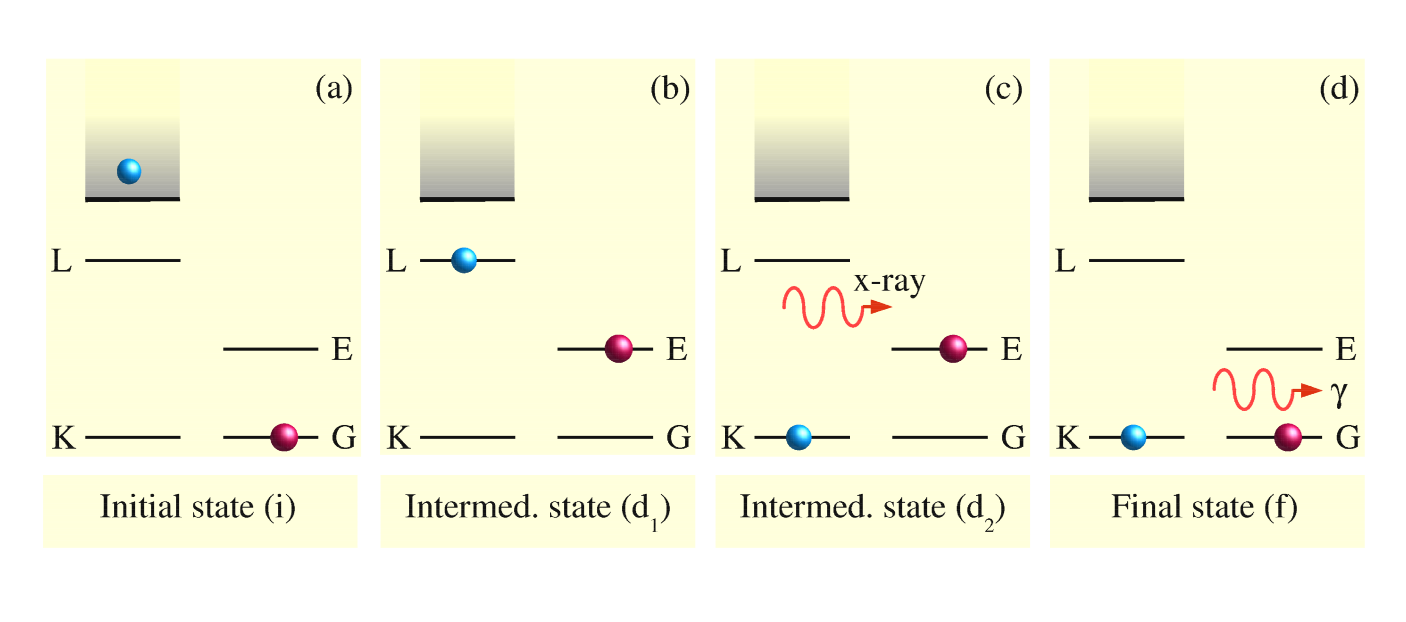}
\caption{\label{figure} Schematic representation of the three-step process of NEECX followed by $\gamma$ decay of the nucleus. A free electron is captured into the $L$ shell of a bare ion with the simultaneous excitation of the nucleus  (a) and (b), followed by the fast x-ray emission from the electronic decay to the ground state (c) and the subsequent radiative de-excitation of the nucleus (d).  The nuclear states are labeled with (G) for the ground state and (E) for the excited state.}
\end{center}
\end{figure*}

 In the resonant process of NEEC, the collision of a highly charged ion with a free electron with matching kinetic energy leads to a resonant capture into an atomic orbital with simultaneous excitation of the nucleus [see illustrations (a) and (b) in Figure~\ref{figure}, picturing NEEC into the $L$ shell of a bare ion]. NEEC is the inverse process of IC and results in an excited nuclear state that can decay either radiatively or by IC. 
First proposed theoretically in Ref.~\cite{Goldanskii}, NEEC is closely related to other processes coupling the nucleus to the atomic shells, such as nuclear excitation by electron transition (NEET) \cite{Morita,Tkalya1}, resonant internal conversion \cite{Karpeshin}  and bound internal conversion (BIC) \cite{Attallah}.   While both NEET  and BIC have been recently confirmed experimentally \cite{Kishimoto,Carreyre}, the experiments aiming at measuring NEEC performed  so far \cite{Izawa,ganil} have failed to observe the effect. 
Most of the theoretical studies giving the magnitude of the NEEC cross sections \cite{Cue1,Cue2,Kimball1,Kimball2,Harston,Palffy1,Palffy2,Palffy3} considered electron recombination occurring into the electronic ground state only. NEEC into an excited electronic state is however an interesting process that leads to rearrangement of the electronic configuration due to subsequent fast x-ray emission  and can thus modify the nuclear decay channels. Furthermore, in certain systems, the fast x-ray transition following NEEC closes the IC decay channel. In this case, NEEC can only be followed by $\gamma$ decay of the nucleus and the nuclear lifetime is then  determined by the $\gamma$ decay rate.

While the possibility of NEEC into excited electronic states has been briefly mentioned in Ref.~\cite{Cue1}, 
 rigorous theoretical calculations taking into account the effect of the x-ray decay are missing. Motivated by this, we investigate in this Letter the concrete enhancement of the population and lifetime of low-lying excited nuclear states that occurs in the process of NEEC followed by the fast decay by x-ray emission of the captured electron. We denote this process by NEECX in accordance with the already established notation for the atomic process of resonant transfer and excitation followed by x-ray emission (RTEX). 
Considering the Experimental Storage Ring (ESR) facility in GSI Darmstadt, we put forward how NEECX in the bare $^{238}_{92}\mathrm{U}$ and $^{232}_{90}\mathrm{Th}$ ions provides a realistic scenario for the observation of the NEEC effect. Beyond the immediate realizability of related laboratory experiments, we argue that this theoretically predicted effect of NEECX on  nuclear excited level population should be taken into account for properly describing high-electron density astrophysical plasmas.

If the electron capture takes place into an excited electronic state, NEEC leads to a doubly-excited intermediate state $d_1$, with both the electronic shell and the nucleus excited [see Figure~\ref{figure} (a) and (b) that picture the capture into the $L$ shell of initially bare ions]. This intermediate state can decay via emission of atomic x-ray or nuclear $\gamma$ photons, or,  alternatively,  nuclear de-excitation by IC may occur. If the x-ray emission is faster than the nuclear decay, the captured electron eventually decays radiatively to the ground state, leading to a second intermediate state $d_2$ [see Figure~\ref{figure} (c)]. For cases in which internal conversion from the $K$ shell is energetically forbidden, the only remaining nuclear decay channel is the radiative one. Thus the integrated cross section for the three-step process of NEECX followed by $\gamma$ decay of the nucleus is substantially increased.

For the cases of the  $^{238}_{92}\mathrm{U}$ and $^{232}_{90}\mathrm{Th}$  actinide nuclei, we
explain the NEECX scenario in detail. Both $^{238}_{92}\mathrm{U}$ and $^{232}_{90}\mathrm{Th}$ have low-lying $2^+$ excited nuclear states, with  energies of $E_{\gamma}=44.910$~keV and 49.369~keV \cite{toi}, respectively, that are connected to the $0^+$ nuclear ground state via an $E2$ transition.
Due to the large binding energy of the $K$-shell electron, \mbox{-131.815~keV} for ${\rm U}^{91+}$ and \mbox{-125.248~keV} for ${\rm Th}^{89+}$ \cite{Soff,MPS}, the decay of the first excited nuclear state by IC of the $1s$ electrons is forbidden.  
We consider NEECX involving  the capture
of a free electron into the $L$ shell of the bare 
ion with the simultaneous excitation of the nucleus into the
 first excited nuclear state, as depicted in Figure \ref{figure}. Electron capture into bound orbitals belonging to higher atomic shells is also possible and can be described in the same manner. 
The electronic excited state decays several orders of magnitude faster than the low-lying nuclear states, e.g. 
the lifetimes of the $L$-shell one-electron configurations of the ${\rm U}^{91+}$ ion are  $5\times 10^{-15}$~s, $2\times 10^{-17}$~s and $3\times 10^{-17}$~s for  the $2s$, $2p_{1/2}$ and $2p_{3/2}$ orbitals, respectively.  For comparison, the nuclear excited state lives approximately $2\times 10^{-10}$~s in the case of neutral atoms and about $10^{-7}$~s in the case of bare ions (see Table~\ref{lifetimes}). Similar numbers apply for ${\rm Th}^{89+}$. The $L$-shell electron will therefore de-excite rapidly to the $K$ shell by emitting a photon, leading to an intermediate state  $d_2$
characterized by the electron in the $K$ shell and the 
nucleus in an excited state. As the IC decay channel is forbidden, de-excitation of the nucleus occurs radiatively, and the 
lifetime of the nuclear excited state is given in this case by the $\gamma$ decay rate, $\tau_{\gamma}=A_{\gamma}^{-1}$. 
For the studied nuclei the $\gamma$ decay of the first excited nuclear state is a much slower process than IC, so that the nuclear lifetime is increased considerably.

In order to describe the three-step process of NEECX together with the subsequent  $\gamma$ decay of the nucleus, we use a Feshbach projection operator
formalism to account for the possible intermediate states,
extending the method presented in Ref.~\cite{Palffy1}. We  can
write the total resonance strength (i.e., the total cross section integrated over
the continuum electron energy) of the process from the initial
state $i$ characterized by the continuum electron and the nucleus
in the ground state to the final state $f$ with the bound electron
and nucleus in their respective ground states via the two intermediate states
$d_1$ and $d_2$ as (in a.u.)
\begin{equation}
S_{\rm NEECX}^{i\to
f}=\frac{2\pi^2}{p^2}\frac{A_{\gamma}^{d_2\to
f}}{\Gamma_{d_2}}\frac{A_{\mathrm{x\, ray}}^{d_1\to
d_2}}{\Gamma_{d_1}} Y^{i\to d_1}_{n}\, .
\end{equation}
Here, $p$ denotes the momentum of the continuum electron,
$A_{\gamma}$ and $A_{\mathrm{x\, ray}}$ are the nuclear and
electronic radiative decay rates, respectively, and $Y^{i\to
d_1}_{n}$ is the NEEC rate \cite{Palffy1}. The width $\Gamma_{d_1}$ of the doubly-excited state
$d_1$ is given by the sum of the nuclear and electronic 
widths, $\Gamma_{d_1}=\Gamma_{\gamma}+\Gamma_{IC}+
\Gamma_{\mathrm{x\, ray}}$, and can be approximated as
$\Gamma_{d_1}\simeq\Gamma_{\mathrm{x\, ray}}$ due to the difference
of magnitude of the electronic and nuclear widths. The
radiative width $\Gamma_{\gamma}$ of the nuclear excited state  determines the width of the second intermediate state $\Gamma_{d_2}$. An additional second term standing for the process in which the
nuclear decay occurs prior to the electronic decay can be neglected due to the long lifetime of the nuclear
excited state. The
electronic widths  and transition rates are calculated with the
OSCL92 module of the GRASP92 package \cite{grasp92}. In the case of NEEC into the
$2p_{3/2}$ and $2s$ orbitals of the bare $^{238}_{92}\mathrm{U}$ and $^{232}_{90}\mathrm{Th}$ ions, we consider the direct
electronic transition to the $1s$ ground state, neglecting the much slower decay channels via the intermediate $2p_{1/2}$ or $2s$ states. 
In Table \ref{res_str} we present the resonance strengths involving
NEECX into the excited $2s$, $2p_{1/2}$ and $2p_{3/2}$ atomic states of bare $\mathrm{U}^{92+}$ and 
$\mathrm{Th}^{90+}$ with subsequent x-ray decay to the $1s$ atomic ground state. For comparison, resonance strengths for NEEC into ions with occupied $K$ shell or $L$ subshells are given, where the electronic capture occurs directly into the ground state without subsequent x-ray emission. 
  Since both $^{238}_{92}\mathrm{U}$ and $^{232}_{90}\mathrm{Th}$ have high atomic numbers $Z$, the different charge states  considered have a small effect on the  NEEC rates involved \cite{Palffy1}. For both NEEC and NEECX, the calculated resonance strengths in Table~\ref{res_str} include the subsequent nuclear decay via $\gamma$ emission.

The half-lives for the $E_{\gamma}=44.910$~keV level of
$^{238}_{92}\mathrm{U}$ and the $E_{\gamma}=49.369$~keV of $^{232}_{90}\mathrm{Th}$ are given as 203~ps and  345~ps, respectively, in Ref.~\cite{toi}. These values correspond to IC coefficients of the neutral atoms  of $\alpha=609$ and $\alpha=327$, respectively \cite{Raman}. 
The IC coefficient is defined 
as the ratio of the IC and $\gamma$ decay rate, $\alpha=A_{\mathrm{IC}}/A_{\gamma}$.
The corresponding $\gamma$ rates do not depend on the electronic
configuration and are  $A_{\gamma}=5.40\times
10^6$~s$^{-1}$ for $^{238}_{92}\mathrm{U}$ and $A_{\gamma}=6.65\times
10^6$~s$^{-1}$ for $^{232}_{90}\mathrm{Th}$. In  ${\rm U}^{91+}$, an electron occupying the $2s$, $2p_{1/2}$ or
$2p_{3/2}$ orbital of the ion accounts already for a partial IC
coefficient of $\alpha=$4, 129 and 67, respectively. Likewise we obtain
$\alpha=$ 2, 69 and 36 for the considered one-electron configurations in  ${\rm Th}^{89+}$.
The lifetimes of the nuclear excited states are therefore prolonged by
more than two orders of magnitude if the captured electron decays to the $K$ shell,  rendering the $L$-shell IC impossible. In Table~\ref{lifetimes} we present values for the
mean lifetimes of the $E_{\gamma}=44.910$~keV level of
$^{238}_{92}\mathrm{U}$ and of the $E_{\gamma}=49.369$~keV level of
$^{232}_{90}\mathrm{Th}$ for neutral atoms \cite{toi} and  our theoretical results for H-like, Li-like, B-like and N-like  ions formed via NEEC or NEECX. 
The third column in Table~\ref{lifetimes} presents the lifetimes of the first excited $^{238}_{92}\mathrm{U}$ and $^{232}_{90}\mathrm{Th}$ states in H-like ions with the electron in the ground state, as occurring after NEECX into bare ions. The nuclear state lifetime is then solely determined by the $\gamma$ decay rate of the excited nuclear state. 
 The nuclear mean lifetimes $\tau_{nl_j}$ for the atomic ground state configurations of Li-like ($1s^22s$), B-like ($1s^22s^22p_{1/2}$) and N-like ($1s^22s^22p^2_{1/2}2p_{3/2}$) ions are given in the fifth column of Table~\ref{lifetimes}. These lifetimes correspond to the final states of NEEC occurring into He-like, Be-like and C-like ions considered in Table~\ref{res_str}. Comparing the third and fifth columns of Table~\ref{lifetimes}, a two orders of magnitude difference in the nuclear lifetime is to be observed for capture into the $2p$ orbitals, depending on whether the  decay to the $K$ shell that suppresses the IC decay channel is possible.
The magnitude of this difference is based on the large value of the IC coefficient $\alpha$, which in turn reflects the  behavior of the $\gamma$ and IC decay rates of the excited nuclear state.

Our theoretical results show that the fast x-ray emission from the electronic decay in NEECX increases both the resonance strength of the process  and the lifetime of the excited nuclear state by up to more than two orders of magnitude. NEECX into the $L$ shell of bare $^{238}_{92}\mathrm{U}$ and $^{232}_{90}\mathrm{Th}$ ions is at present the most promising choice for an experimental verification of NEEC, since the calculated resonance strength values exceed, to the best of our knowledge, the ones of all other considered scenarios. 
 Apart from the large resonance strength values, two other advantages should be noted. Since the electron capture occurs into an excited electronic shell, the width of the capture state is no longer  given by the natural width of the nuclear excited state, but it is dominated by the fast x-ray transition, $\Gamma_{d_1}\simeq\Gamma_{\mathrm{x\, ray}}$. With electronic widths  on the order of tens of eV, orders of magnitude larger than the ones for NEEC into ground state electronic configurations \cite{Palffy2},  the continuum electron energy resonance conditions can be easily fulfilled experimentally. Furthermore, as it will be explained in the next paragraphs, the convenient nuclear lifetime renders possible the separation of the $\gamma$ decay photons  from the background signal 
in a NEEC experiment involving the studied actinide isotopes. 

A major difficulty in observing  NEEC experimentally arises from the  background of atomic photorecombination processes, in particular from the non-resonant channel of radiative recombination (RR).
Due to identical initial and final states of NEEC followed by $\gamma$ decay of the nucleus and RR, and to the dominance of the latter,  it is
practically impossible to distinguish between the two 
processes \cite{Palffy2}. The RR photon yields for typical experimental conditions exceed the ones of the $\gamma$ photons following NEEC by orders of magnitude, resulting in a signal-to-background ratio of less than $10^{-3}$ even for the most promising cases \cite{Palffy2}.
 Recently, a theoretical study of the
angular distribution of the photons emitted in the two processes
showed that in the case of $E2$ nuclear transitions the emission patterns
of the two processes are different \cite{Palffy3}, thus slightly improving the contrast. Nevertheless, angular
sensitive measurements are even more demanding than those
attempting to measure total cross sections. In practice,
the different time scales on which NEEC and RR occur can be
useful for eliminating the background in an experiment. 
In Ref.~\cite{Cue1} this concept has been proposed in the framework of an NEEC experiment that considers an ion beam channeling through a single crystal acting as an electron target. While RR is practically instantaneous and its photon signals last typically for $10^{-14}$~s,  NEEC involves intermediate excited nuclear states which decay after $10^{-10}$~s or even later following the recombination. During this time, the channeling ions traverse a certain distance from the solid target, so that a spatial separation between the RR and NEEC photons is achieved. The authors of Ref.~\cite{Cue1} have proposed the use of a beam-foil geometry to practically eliminate all atomic background events.

  Unfortunately, a source of background that cannot be discriminated by a beam-foil arrangement is the Coulomb excitation of the same projectile nuclear level by the target nuclei.  This can be however avoided if an electron target is used in storage ring experiments. In the ESR at the GSI Darmstadt the ions  cycle with a typical re\-vo\-lu\-tion frequency of  3~MHz and the free electrons are 
provided by  the electron cooler
\cite{Bra02,Bra03}. 
 Similarly to the concept presented in \cite{Cue1}, the
different time scales of RR and NEEC have as a result a
spatial separation of the  photon emissions.  While the RR photons
will be emitted instantaneously in the region of the electron
target, the radiative decay of the nucleus will occur
later, after the ions have already traveled a certain distance in
the storage ring. With the ion velocity being 71$\%$ of the speed of light, 
corresponding to an ion energy that is used at the ESR for stochastic cooling, typical lifetimes of
10-100~ns correspond to 2-20~m distance between the prompt and delayed photon emission. The electron cooler is used as a target for free electrons and can be tuned to match the resonance condition for NEEC. 
Ions recombined in the cooler are separated from the primary beam in the first bending magnet of the storage ring and can be counted with an efficiency close to unity. By choosing isotopes with convenient excited state lifetimes on
the order of 10-100 ns, the spatial separation can be
determined such that a direct observation of the $\gamma$ photons following
NEEC can be performed with almost complete suppression  of the RR
background. 

The case of the actinide nuclei under investigation in this work is therefore especially appealing and illustrative in this respect due to the significantly enhanced yield of delayed $\gamma$-rays. For a future experiment at the ESR in GSI Darmstadt we envisage to position $\gamma$-ray detectors with large detection efficiency (e.g., large NaI scintillators) in  special detector chambers after the electron cooler. The prompt radiation coming from the cooler such as RR or bremsstrahlung will be totally shielded from the detector, so that only delayed photons emitted downstream from the shielding will produce a signal. Solely the signals associated with ions that have captured one electron in the cooler will be recorded in an appropriate time window, so that the number of random coincidences with the natural $\gamma$-ray-background events is negligible. Thus, when scanning  the relative electron-ion energy in several steps over the NEECX resonance, we expect practically no count rate outside the resonance. Due to the small background, the Doppler broadening of the $\gamma$ energy quanta is not expected to play any role in the experiment.
The rate for the occurrence of NEECX with subsequent $\gamma$ decay in the proposed experiment is estimated for maximum beam luminosity and present ESR parameters \cite{Steck} to be 4.6 events per minute, or 273 events per hour. The count rate at the resonance depends very strongly on the detection efficiency and can be as low as a
few counts per day. This is a serious challenge that could be tackled by a
careful optimization of the detection efficiency of a multiple-detector
array along the beam line.

Finally, we discuss the role of NEECX as recombination mechanism in astrophysical plasma environments with high electron densities.
In the atomic physics counterpart of NEEC---the process of dielectronic recombination (DR)---a free electron gets captured into an ion with the simultaneous excitation of a bound electron, leading to a doubly-excited electronic state \cite{Massey}. Owing to recent improvements in sensitivity and resolution of astronomical observations, a detailed understanding of DR becomes increasingly important for the interpretation of the high-quality spectral data \cite{Eike,Daniel}. Since DR and NEEC followed by the $\gamma$ decay of the nucleus differ only in the excitation channel, one may speculate about the potential impact of NEEC in the field of astrophysics.   Although the resonance strengths for NEEC followed by $\gamma$ decay of the nucleus are a
few orders of magnitude smaller than the ones for DR, the
NEEC process is also expected to occur in astrophysical plasmas for heavy ions with suitable nuclear excitations. Particularly at large electron densities prevailing in hot dense plasmas in nucleosynthesis scenarios, nuclear mechanisms coupling to the atomic shell such as NEEC, NEET, IC, and BIC are predicted to be of relevance \cite{Morel}.  NEECX can occur also into higher atomic shells of highly charged ions, with subsequent slow $\gamma$ decay of the  excited nucleus. Similarly to DR \cite{Bra02,Bra03}, a whole series of NEECX resonances spans the collision energy range up to the series limit at the nuclear excitation energy. The formation of nuclear excited states via NEECX depends decisively on the environment temperature and ionization stage. Especially for certain key isotopes  used as cosmochronometers or nuclear thermometers in astrophysical plasmas \cite{forty_years,Kaeppeler,Klay} that provide important information about the age of the universe and the temperature in nucleosynthesis processes,  plasma models should be refined to account for coupling of nuclei to the atomic shells.  In nucleosynthesis scenarios, NEEC with or without subsequent x-ray decay may play an important role via effective population of nuclear excited states or in triggering the  decay of isomers by exciting the nucleus  to an intermediate level that subsequently decays directly or in cascades to the ground state. For intermediate levels at less than 200~keV above the isomer, NEEC provides an efficient branching ratio for releasing the energy of the metastable state \cite{PEK}. Via the time-reversed process, low-lying isomeric states can also be populated by nuclear excitation mechanisms such as NEEC or NEECX. Apart from the significance of these processes in astrophysical plasmas, a number of potential applications related to the controlled release of nuclear energy on demand  have been suggested, such as nuclear batteries \cite{Walker,nat_phys}.

In conclusion, our theoretical calculations show that the fast electronic x-ray decay following the resonant capture  leads to a substantial increase of the nuclear excited state population and the resonance strength of NEEC followed by $\gamma$ decay of the nucleus.
 The enhancement of the  $\gamma$ photon yield and the convenient nuclear lifetimes of the heavy $^{238}_{92}\mathrm{U}$ and $^{232}_{90}\mathrm{Th}$ actinides can be particularly useful for the possible experimental observation of NEEC. As a mechanism for preparation of nuclear excited species by electron recombination, NEECX may also have relevance in high electron-density plasmas.

%%%%%%%%%%%%%%%%%%%%%%%%%%%%%%%%%%%%%%%%%%%%%%%%%%%%%%%%%%%%%%%%%%%%%%%%

\bibliography{palffy_arxiv}

\begin{table}[h]
\caption{\label{res_str} Continuum electron resonance energies $E_c$ and resonance strengths for NEEC and NEECX, both followed by $\gamma$  decay of the nucleus for $^{238}_{92}\mathrm{U}$ and $^{232}_{90}\mathrm{Th}$. In the fourth column, $S_{\rm NEEC}$ is the resonance strength for NEEC  into the $2s_{1/2}$ orbital of initially He-like ions, $2p_{1/2}$ orbital of Be-like ions and $2p_{3/2}$ orbital of C-like ions, respectively, calculated following the formalism described  in Ref.~\cite{Palffy1}. $S_{\rm NEECX}$ stands for the resonance strength of NEECX into bare ions. The capture orbital is denoted by $nl_j$.}
\center{
\begin{tabular}{lccccc}
\hline
& & \multicolumn{2}{c}{He-, Be-, C-like ions} & \multicolumn{2}{c}{bare ions}\\
$^A_Z\mathrm{X}$ &$nl_j$ & $E_c$ (keV)& $S_{\rm NEEC}$(b eV)& $E_c$ (keV) & $S_{\rm NEECX}$(b
eV) \\ \hline
            & $2s_{1/2}$ & 12.072 & $8.8\times 10^{-3}$ &10.783 &$5.4\times 10^{-2}$ \\
$^{238}_{92}\mathrm{U}$   &$2p_{1/2}$ & 13.262 & $9.2\times 10^{-3}$ &10.706 &
1.58\\
            & $2p_{3/2}$ & 18.329 & $2.7\times 10^{-3}$ &15.269 & 1.16 \\
\hline
            & $2s_{1/2}$ & 18.244 & $5.9\times 10^{-3}$  & 17.001 &  $2.0\times 10^{-2}$ \\
$^{232}_{90}\mathrm{Th}$&  $2p_{1/2}$ &19.400 & $7.7\times 10^{-3}$  & 16.935 &
$6.5\times 10^{-1}$ \\
            & $2p_{3/2}$ & 24.017 & $2.6\times 10^{-3}$ &21.040 & $5.5\times 10^{-1}$\\
\hline
\end{tabular}}

\end{table}
\begin{table}
\caption{\label{lifetimes} Mean-lives $\tau$ for the first nuclear
excited states of $^{238}_{92}\mathrm{U}$ and
$^{232}_{90}\mathrm{Th}$. In the second column,
$\tau_{0}$ stands for the mean lifetime
corresponding to the nucleus of a neutral atom, while  the third
column contains the values $\tau_{\gamma}$ for the bare ion
or H-like ion with the electron in the ground state. 
Nuclear mean-lives of Li-like ($2s_{1/2}$), B-like ($2p_{1/2}$) and N-like ($2p_{3/2}$) ions in their electronic ground states are presented in the last column.}
\center{
\begin{tabular}{lc@{$\quad$}c@{$\quad$}c@{$\quad$}c}
\hline
& neutral & bare/H-like & & Li-, B-, N-like\\
$^A_Z\mathrm{X}$&$\tau_{0}$ (ns) &$\tau_{\gamma}$ (ns)  & $nl_j$ &$\tau_{nl_j}$ (ns)\\
  \hline
            &     &     &$2s_{1/2}$  &36.2\\
$^{238}_{92}\mathrm{U}$ & 0.292 & 185 & $2p_{1/2}$ &1.54 \\
            &     &     &   $2p_{3/2}$  &0.67\\
\hline
            &     &  & $2s_{1/2}$ &50.1\\
$^{232}_{90}\mathrm{Th}$& 0.497 & 150 & $2p_{1/2}$ &2.35\\
            &     &  &  $2p_{3/2}$ &1.04\\
\hline
\end{tabular}}

\end{table}

\end{document}